# Plasmon recombination in narrowgap HgTe quantum wells

**V Ya Aleshkin[1,2], G Alymov[3], A A Dubinov[1,2], V I Gavrilenko[1,2] and F Teppe[4]**

[1] Department of Semiconductor Physics, Institute for Physics of Microstructures RAS, Nizhny Novgorod, 603950, Russia
[2] Lobachevsky State University of Nizhny Novgorod, Nizhny Novgorod, 603950, Russia
[3] Laboratory of 2d Materials for Optoelectronics, Moscow Institute of Physics and Technology, Dolgoprudny 141700, Russia
[4] Laboratoire Charles Coulomb, Université de Montpellier, Centre National de la Recherche Scientique, 34095 Montpellier, France

E-mail: aleshkin@ipmras.ru

**Abstract**

The dispersion laws of two-dimensional plasmons in narrow-gap HgTe/CdHgTe quantum wells are calculated taking into account the spatial dispersion of the electron susceptibility. At the energy scale of the band gap the dependence of plasmon frequencies on the wave vector is shown to be close to linear that changes significantly the critical concentration of noneqilibrium electron-hole gas corresponding to "switching-on" the carrier recombination with plasmon emission. The recombination rates with the plasmon emission have been calculated. The , "plasmon" recombination is shown to dominate at the carrier concentration over $(1.2\text{-}2)\ 10^{11}\ cm^{-2}$ in a 5-nm-wide HgTe quantum well (band gap of 35 meV) that makes plasmon generation (spasing) in THz frequency range feasible.

## 1. Introduction

Despite immeasurable bulk of work conducted recently, terahertz (THz) frequency range still lacks compact and effective emitters that are required for many applications. Quantum cascade lasers (QCLs) demonstrate remarkable performance in the spectral range from 1 THz to 5 THz [1, 2] and above 15 THz [3]. Nowadays, the majority of the QCLs exploit GaAs and InP semiconductors, in which lattice absorption becomes very strong at wavelengths longer than 20 μm. In 20–28 μm range, operation of the QCLs has been demonstrated only for several specific wavelengths [4-7]. GaN QCLs attempt to enter the 5–15 THz "gap" from the low-frequency side, but their performance is yet to be improved [1, 8]. The spectral range from 28 μm to 60 μm is now partly covered only with the lead salt diode lasers, which provide emission wavelengths up to 50 μm [9]. In PbSnSe(Te) ternary alloys the nonadiative Auger recombination is known to be suppressed due to the symmetry between the dispersion laws in the conduction band and in the valence band [10]. For symmetric electron-hole dispersion energy-momentum conservation laws impose an energetic threshold for Auger processes. However, the output power and operation temperature of lead salt lasers are limited by the growth technology.

In recent years, a significant advance in regard to the wavelength of coherent emission has been demonstrated in Hg(Cd)Te/CdHgTe quantum well (QW) heterostructures, grown by molecular beam epitaxy (MBE) (up to 19,5 μm [11] and 24 μm [12]). The abovementioned symmetry of electron-hole dispersion laws takes place in such QWs as well, whereas transverse optical phonons in HgCdTe have low energies compared to GaAs and InP, suggesting that lasing should be possible at wavelengths up to 50 um (6.2 THz) [13]. However, quite strong two-phonon absorption of 40-100 cm-1 is not so easy to overcome in the wavelength range above 35 μm (see, e.g. [14]). Moreover, growing thick (over 20 μm) HgCdTe structures, required to form a dielectric waveguide for such long radiation wavelengths, is a challenge for MBE technology.

Employing two-dimensional (2D) /surface plasmon amplification in QWs one can achieve a dramatic increase in the gain in interband transitions. Due to the relatively small group velocity of the plasmon, the amplification coefficient can be large, substantially exceeding the amplification coefficient in structures with dielectric light waveguides. During the last decade 2D plasmons have attracted a lot of attention in regard to generating and detecting far-infrared radiation [15-19]. In graphene and narrow-gap QWs 2D plasmons can be generated at the interband transitions of electrons [17-20], i.e. plasmon emission is an "additional" mechanism of the interband electron recombination in such systems. The electric field of 2D plasma wave is highly localized near the QW, so there is no need to grow thick widegap layers for dielectric waveguide. However, while 2D plasmons and their role in recombination in graphene have been studied in details (see, e.g. [20]), there are only a few works on plasmons in narrow-gap HgTe QWs [13, 19]. Note that both in Ref.13 and Ref.19 the spatial dispersion of the electron-hole gas susceptibility was not taken into account. As a result, these works implemented the square root dependence of the plasmon frequency $\omega$ on the plasmon wavevector $q$ $\omega \sim \sqrt{q}$, which is a poor approximation for the plasmon energies larger than 20 meV, as shown in the present work.

This paper is organized as follows. In Section 2 we find the dispersion law for 2D plasmons in narrow-gap HgTe QWs focusing on plasmon energies of 30-60 meV. Section 3 is devoted to investigation of the "plasmon" contribution into the interband recombination of nonequilibrium carriers in HgTe QWs.

## 2. Plasmon in HgTe quantum well structure

We consider the plasmon dispersion in the quasistatic approximation, when the plasmon wavevector $q$ is much larger than that of the photon with the same frequency. The expression for 2D dimensional plasmon dispersion is well known for the case when the spatial dispersion of the the electron-hole gas susceptibility is neglected [21]:

$$\omega = \sqrt{\frac{2\pi e^2 q}{\kappa}\left(\frac{n}{m_e} + \frac{p}{m_h}\right)}. \qquad (1)$$

In (1) $n$ and $p$ are sheet electron and hole concentrations respectively, $\kappa$ is the dielectric constant of the medium at the plasmon frequency, $m_e$ and $m_h$ are electron and hole effective mass respectively, and $e$ is the electron charge. The spatial dispersion of the susceptibility of the electron-hole gas can be disregarded for plasmons with small wavevectors. However, as shown below, this is not the case for the plasmons with energies exceeding the band gap of HgTe QWs under consideration. In what follows we assume that the QW thickness is small compared to $1/q$, which is a good approximation for the QWs under study (see, e.g. [19]). In this case, the dispersion law of 2D plasmons $\omega(q)$ that takes into account the spatial dispersion can be obtained from the following equation:

$$1 + \frac{2\pi q}{\kappa}\big(\chi_e(q,\omega) + \chi_h(q,\omega)\big) = 0 \qquad (2)$$

[21] where $\chi_{e,h}(q,\omega)$ are the susceptibilities of 2D electron ($e$) and hole ($h$) gases:

$$\chi_{e,h}(q,\omega) = \frac{2e^2}{q^2(2\pi)^2}\int d^2k \left[\frac{f_{e,h}(\mathbf{k}) - f_{e,h}(\mathbf{k+q})}{\varepsilon_{e,h}(\mathbf{k+q}) - \varepsilon_{e,h}(\mathbf{k}) - \hbar\omega - i\hbar v_{e,h}}\right] \qquad (3)$$

In equation (3) $\varepsilon_{e,h}(\mathbf{k})$ are electron and hole energies, $f_{e,h}(\mathbf{k})$ are electron and hole distribution functions, $\hbar$ is the Planck constant. We note that equation (2) takes into account only the intraband contributions of electrons and holes to the susceptibility. Numerical estimations show that the contribution of the interband electronic transitions to the susceptibility is negligible. Equation (3) can be obtained from the expression for the density matrix in the τ-approximation for the relaxation of the off-diagonal components of the density matrix [22]. Frequencies $v_{e,h}$, which stand for the electron and hole phase relaxation, are further assumed to be independent of the momentum for the sake of simplicity. They are set equal to the momentum relaxation rates that can be estimated from the measured mobilities of electron and hole gases $\mu_{e,h}$ in similar QWs:

$$\mu_{e,h} = e/(v_{e,h}m_{e,h}),$$

where $m_{e,h}$ is electron or hole effective mass at the corresponding Fermi energy. Note that in equation (3) the wavevector $q$ is assumed to be real, while the frequency $\omega$ can be complex.

In the calculations, the distribution function was chosen as:

$$f_e(\mathbf{k}) = \left(1 + \exp\left(\frac{\varepsilon_e(\mathbf{k}) - F_c}{k_B T^*}\right)\right)^{-1},$$

$$f_h(\mathbf{k}) = \left(1 + \exp\left(\frac{\varepsilon_h(\mathbf{k}) + F_v}{k_B T^*}\right)\right)^{-1}$$

where $k_B$ is the Boltzmann constant, $T^*$ is the effective carrier temperature, $F_{c,v}$ are the quasi Fermi level in the conduction band and valence band respectively.

As can be seen from equation (2) and equation (3), it is necessary to calculate the dispersion laws of electrons and holes to find the dispersion law of the 2D plasmon. To this end, we used the 4-band Kane Hamiltonian [23] taking into account the deformation effects. Calculation details and material parameters can be found in Ref. [24].

In the following we present the calculation results for a 5-nm-wide typical HgTe/Cd$_{0.7}$Hg$_{0.3}$Te QW grown on GaAs substrate with CdTe(013) buffer layer (see e.g. [25]). The band gap in this QW is about 35 meV. The calculated band diagram is given in the figure 1.

Figure 2 shows the calculated dependences of the real part of the plasmon frequency on its wave vector $\omega(q)$ for the same QW for the lattice temperature $T = 4.2$ K and various nonequilibrium carrier concentrations ($n = p$) and effective carrier temperatures $T^*$. It can be clearly seen from this figure that $\omega$ is proportional to $q^{1/2}$ at small $q$ ($q < 0.02$ nm$^{-1}$), while at larger $q$ the dependence $\omega(q)$ is close to the linear one. The reason for the deviation from the square root dependence is the spatial dispersion of the susceptibility (see equation (3)). Note that at large wavevectors the dependence of the plasmon frequency on the wavevector is close to the linear one in graphene as well [20]. Figure 2 shows a weak dependence of the 2D plasmon dispersion on the temperature of charge carriers.

To illustrate how the spatial dispersion of susceptibility affects the 2D plasmon dispersion Figure 2 also shows two $\omega(q)$ dependences calculated neglecting the spatial dispersion (see two dark yellow curves, $n = p = 4*10^{11}$ cm$^{-2}$). The upper one (0) was calculated by substituting the electron and hole effective masses at the bottom of the conduction band and at the top of the valence band, correspondingly into equation (1), the same way as in the work [19]. However, within this approach it seems more reasonable to substitute the conductivity effective masses of the charge carriers at the Fermi energy $m_{e,h} = 2\varepsilon\hbar^2(\partial\varepsilon_{e,h}/\partial k)^{-2}\big|_{\varepsilon=E_F}$, which are significantly larger due to the band nonparabolicity. Exploiting masses at the Fermi energy results in the bottom dependence (F). It goes significantly lower than the curve that was calculated taking into account the spatial dispersion of the susceptibility for the same carrier concentration $n = p = 4*10^{11}$ cm$^{-2}$ (red curve in the figure 2) and does not cross $E_g^{eff}(q)$ dependence (green curve in figure 2) at reasonable wavevectors.

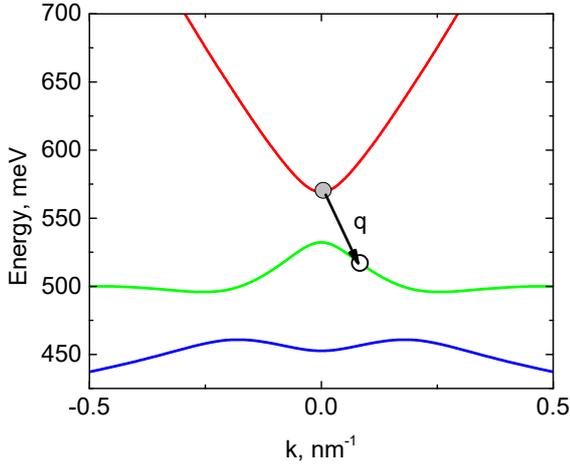

**Figure 1.** Calculated band diagram of a 5-nm-wide HgTe/Cd$_{0.7}$Hg$_{0.3}$Te(013) QW at $T = 4.2$ K. The wave vector **k** is in [100] direction. The arrow illustrates the interband recombination with a plasmon emission.

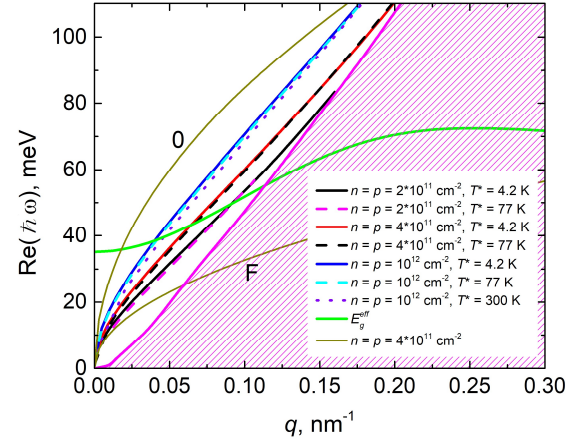

**Figure 2.** Calculated dependences of the real part of plasmon energy $\hbar\omega$ on the wavevector (**q** ‖ [100]) for a 5-nm-wide HgTe/Cd$_{0.7}$Hg$_{0.3}$Te(013) QW with different carrier concentrations $n = p$ and effective carrier temperatures $T^*$. The green line shows the dependence of the minimum energy of the interband transition due to plasmon emission $E_g^{eff}(q)$ (i.e. the "effective" bandgap) on the plasmon wavevector. Two dark yellow curves correspond to the square root dependencies of the plasmon energy on the wavevector given by equation (1) (i.e. the spatial dispersion of susceptibility in not taken into account). The upper one (0) has been calculated for electron and hole effective masses at the corresponding band edges while the bottom one (F) – for those at the Fermi energies. The hatched area correspond to the Landau damping calculated for $T = T^* = 4.2$ K.

The hatched area in figure 2 corresponds to the region where plasmons are poorly defined on account of Landau damping that delivers a large attenuation of the plasmon, exceeding possible amplification due to the interband electron transitions [19]. The upper boundary of this area corresponds to the plasmon absorption by electrons with the Fermi energy, i.e. it is found from the equation $\hbar\omega_{max}(q) = \varepsilon(k_F + q) - \varepsilon(k_F)$ [19]. Since the Fermi velocity of holes is less than that of electrons at $n = p$, the plasmon absorption by holes has no effect on $\hbar\omega_{max}(q)$. Note that with the increase in the effective carrier temperature $\hbar\omega_{max}(q)$ is enhanced owing to the filling by electrons states above the Fermi energy.

Due to the anisotropy of both the electron and the hole dispersion the plasmon spectrum is also anisotropic; however, this anisotropy is rather weak. To illustrate it, Figure 3 shows the dependence of the plasmon energy vs. the direction of the wavevector with a fixed module $q = 0.1$ nm$^{-1}$. One can see that the plasmon energy gets maximum at **q** ‖ [03-1] that makes the angle $\pi/2$ with [100] direction. Note that the difference between the maximum and the minimum energies is only about 2%.

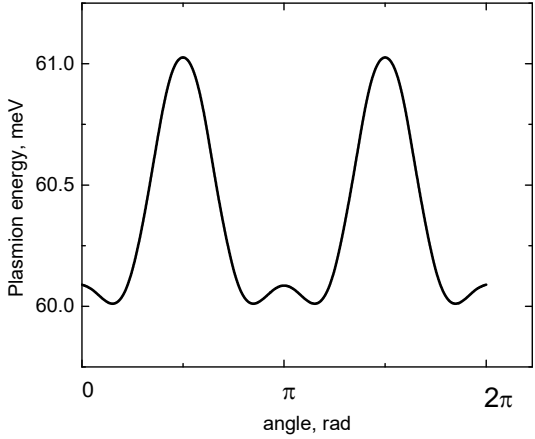

**Figure 3.** Plasmon energy versus propagation angle (calculated from [100] direction) for $q = 0.1$ nm$^{-1}$, $n = p = 2*10^{11}$ cm$^{-2}$, $T = T^* = 4.2$ K.

In the conclusion of this section it is worth considering the dependences of real and imaginary parts of the plasmon frequency on $v_{e,h}$. Within a wide range of $v_{e,h}$ corresponding to the experimentally observed mobilities [25,26] ($\mu_e > 100$ cm$^2$/V*s, $\mu_h > 10$ cm$^2$/V*s), the real part of the frequency is practically independent of $v_{e,h}$. The imaginary part is of the same order as $v_{e,h}$, see figure 4.

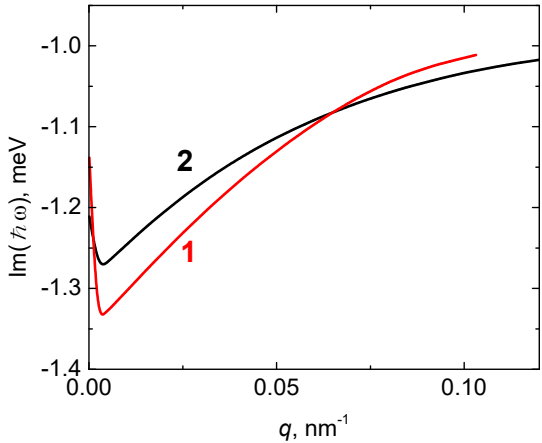

**Figure 4.** Calculated dependences of the imaginary part of the plasmon energy on the wave vector for $\hbar v_e = 1$ meV, $\hbar v_h = 2$ meV. $T = T^* = 4.2$ K, $n = p$, cm$^{-2}$: $10^{11}$ (1), $2*10^{11}$ (2).

## 3. Plasmon recombination mechanism in narrow-gap quantum wells

Plasmons with energies exceeding the effective bandgap can participate in the recombination of the nonequilibrium carriers. To find the recombination rate associated with the emission of a plasmon, it is necessary to quantize the plasmon field. One should keep in mind that the plasmon field energy consists of two parts: the electromagnetic part and the kinetic one. In the quasistatic approximation, the electromagnetic part of the plasmon energy is zero. Carrying out the standard quantization procedure, one can obtain the following expression for the vector potential operator lying in the plane of the quantum well:

$$\mathbf{A}_{\|} = -c\sum_{\mathbf{q}} \frac{\mathbf{q}}{\omega} \sqrt{\frac{2\hbar}{Sq^2}\left(\frac{\partial \chi}{\partial \omega}\right)^{-1}} \left(c_{\mathbf{k}}\exp(i\mathbf{q}\mathbf{r}-i\omega t) + c_{\mathbf{k}}^*\exp(-i\mathbf{q}\mathbf{r}+i\omega t)\right) \quad (4)$$

where $c$ is the speed of light in the vacuum, $S$ is the area of the quantum well, $c_{\mathbf{k}}^*$, $c_{\mathbf{k}}$ are the plasmon creation and annihilation operators, $\chi = \chi_e(q,\omega) + \chi_h(q,\omega)$. Using Equation (4) and the golden rule of quantum mechanics it is possible to obtain the following expression for the probability of spontaneous recombination of an electron with the wavevector k due to the emission of a 2D plasmon:

$$W(\mathbf{k}) = \frac{e^2}{2\pi}\int d^2q \frac{|\mathbf{v}_{if}\mathbf{q}|^2}{q^2\omega^2}\left(\frac{\partial \chi(q,\omega)}{\partial \omega}\right)^{-1}\delta\left(\varepsilon_e(\mathbf{k}) + \varepsilon_h(\mathbf{k}-\mathbf{q}) - \hbar\omega(\mathbf{q})\right)f_h(\mathbf{k}-\mathbf{q}) \quad (5)$$

where $\mathbf{v}_{i,f}$ is the matrix element of the velocity operator between the initial state of the electron in the conduction band and its final state in the valence band. The recombination rate due to this mechanism is:

$$R = \frac{2}{(2\pi)^2}\int W(\mathbf{k})f_e(\mathbf{k})d^2k \quad (6)$$

As it has been shown in the previous section, the dependence of the plasmon energy on the direction of its propagation is weak. Therefore, when calculating the rate of spontaneous recombination, we assume that the plasmon frequency and the susceptibility of the electron gas are independent of the direction of the plasmon wavevector. In this approximation, Eq. (6) can be represented as:

$$R = \frac{2}{(2\pi)^2}\int d^2k d^2q \frac{|\mathbf{v}_{if}\mathbf{q}|^2}{q^2\omega^2}\left(\frac{\partial \chi(q,\omega)}{\partial \omega}\right)^{-1}\delta\left(\varepsilon_e(\mathbf{k}) + \varepsilon_h(\mathbf{k}-\mathbf{q}) - \hbar\omega(q)\right)f_h(\mathbf{k}-\mathbf{q})f_e(k) \quad (7)$$

Integration over $q$ in equation (7) is carried out in the interval $0 \leq q \leq q_{\max}$, where $q_{\max}$ corresponds to the intersection of the Re($\hbar\omega(q)$) dependence with the hatched area in the figure 2 where plasmons are poorly defined due to Landau damping. It can be seen from equation (7) that, in contrast to recombination with the emission of photons, one cannot neglect the change in the electron wave vector in the case of recombination with the emission of plasmons, i.e. transitions with plasmon emission are not vertical (see figure 1). To comply with the energy and momentum conservation laws at such transitions, the plasmon energy with the wavevector $q$ must exceed the effective bandgap $E_g^{\text{eff}}(q)$ (green curve in the figure 2) that is the minimum of the function $\varepsilon_e(\mathbf{k}) + \varepsilon_h(\mathbf{k}-\mathbf{q})$ in the variable **k**. The latter is possible when the carrier concentration exceeds a certain threshold value [19]. Figure 2 shows that as the carrier concentration is decreased, the intersection of the dependences $E_g^{\text{eff}}(q)$ and Re($\hbar\omega(q)$) occurs at higher plasmon energies, exceeding significantly the actual bandgap. Thus, the lower the carrier concentration gets, the higher is the energy of the plasmon that can take part in the recombination. Therefore, only carriers with kinetic energy high enough can participate in the plasmon recombination, while the number of high-energy carriers decreases with decreasing concentration. Thus, the probability of the plasmon recombination drops with a decrease in the carrier concentration. Moreover, we should also exclude from the consideration those plasmons, for which the dispersion $\text{Re}(\hbar\omega(q))$ crosses the $E_g^{\text{eff}}(q)$ dependence in the hatched area of the figure 2 (where plasmons are poorly defined because of Landau damping). Thus, below a certain critical carrier concentration the plasmon generation is "switched-off". For 5-nm-wide HgTe QW under consideration at $T = T^* = 4.2$ K this critical concentration is $1.2*10^{11}$ cm$^{-2}$. Note that in Ref.19, where the spatial dispersion of the susceptibility of charge carriers has not been taken into account, the critical concentration was estimated as $2*10^{11}$ cm$^{-2}$. The critical concentration increases with the effective temperature, because the area of Landau damping enlarges.

Dependences of the recombination frequencies $1/\tau = R/n$ on the effective carrier temperature for two concentrations of the nonequilibrium carriers are given in the figure 5. One can see that the increase in the effective temperature results in the decrease of the plasmon recombination frequency. The reason is a decrease in the filling of electron and hole states involved in the plasmon recombination with the rise of effective carrier temperature.

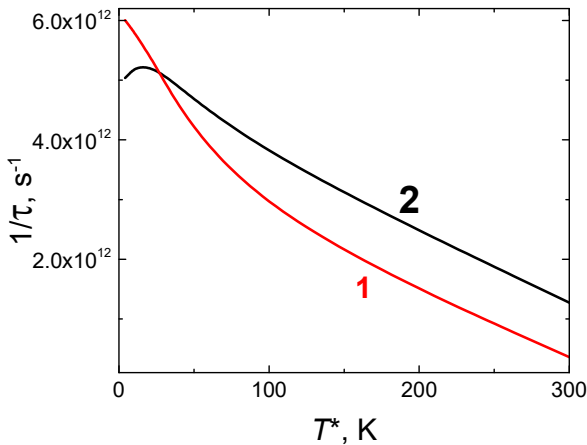

**Figure 5.** Dependences of the plasmon recombination frequency on the effective carrier temperature $T^*$ for two concentrations $n = p$, cm$^{-2}$: Line 1 corresponds to n= $5*10^{11}$, line corresponds to n= $10^{12}$. Lattice temperature is 4.2 K.

It is worth comparing different recombination mechanisms for 5-nm-wide HgTe/Cd$_{0.7}$Hg$_{0.3}$Te QW under consideration. In a QW with low number of defects the Shockley-Read-Hall recombination rate is negligible compared to the radiative recombination one [27]. Figure 6 presents the inverse carrier lifetime versus carrier concentration for three types of recombination: the radiative one, Auger recombination, and the recombination with the emission of plasmons. The inverse carrier lifetimes for radiative recombination and Auger recombination were calculated in the framework of the previously used models that are given in Ref. [27] and [28], respectively. Though nonradiative Auger recombination is supposed to be suppressed in HgTe QW due to a symmetry of the electron-hole energy-momentum laws (see figure 1) its probability increases with reducing the bandgap while that of the radiative recombination, on the contrary, decreases. One can see from figure 6 that for the QW under consideration having the band gap of 35 meV (that corresponds to the wavelength of 35 μm) the probability of Auger recombination is (in most cases) several orders higher than that of the radiative recombination. This complicates furthering the stimulated photon emission above recently achieved wavelength 24 μm [12]. Figure 6 also shows that for carrier concentrations over $(1.2-2)*10^{11}$ cm$^{-2}$, the plasmon recombination is "switched-on" and its frequency is several orders of magnitude higher than that of Auger recombination. This opens a possibility for plasmon generation (spasing) in THz range with subsequent plasmon conversion into the electromagnetic radiation in free space.

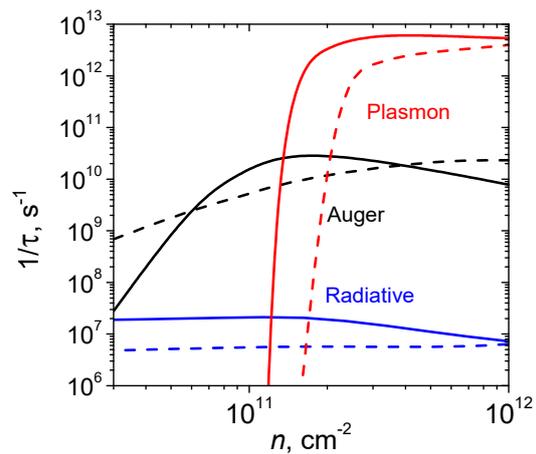

**Figure 6.** Dependences of the recombination frequency (inverse lifetime) on the concentration of nonequilibrium carriers for three recombination mechanisms in 5-nm-wide HgTe QW. Solid lines correspond to the effective carrier temperature 4.2 K, dashed ones – to that of 77 K. Lattice temperature is 4.2 K.

## 4. Conclusion

To conclude, we have calculated the dispersion laws of 2D plasmons with energies exceeding the bandgap in a narrow-gap HgTe quantum well with nonequilibrium electron-hole gas. It is shown that taking into account the spatial dispersion of the charge carriers susceptibility is crucial for the proper description of the plasmon dispersion. At the energy scale of the order of the bandgap the plasmon energy demonstrates nearly linear dependence on the wavevector rather than the square root one as considered earlier. The latter results in significant changes in the critical carrier concentration for the onset of the interband recombination via plasmon emission. The corresponding recombination rates have been calculated showing that the plasmon recombination mechanism dominates in narrow-gap quantum wells at high enough concentrations of nonequilibrium carriers ($n > 1.2*10^{11}$ cm$^{-2}$ for 5-nm-wide HgTe QW). Thus, the plasmon recombination with subsequent plasmon conversion into the outcoming photons can be expoited for developing highly effective emitters in THz range, namely in $5 - 10$ THz region, where QCLs are scarce at the moment.

**Acknowledgements**

This work was supported by the Russian Science Foundation (RSF-ANR Grant # 20-42-09039) and by the French Agence Nationale pour la Recherche (Colector project). The authors are grateful to V.V. Rumyantsev for helpful discussions.

**Appendix 1**
**Formula for susceptibility of the electron gas in a quantum well**
Let us consider the effect of an electric field of an electromagnetic wave on the electron gas. Let the electric field present in the sample create an electric potential

$$\varphi(\mathbf{r},t) = \varphi_0 \exp(i\mathbf{q}\mathbf{r} - i\omega t) \qquad (1.1)$$

This potential leads to the appearance of an additive to the potential energy of the form

$$U(\mathbf{r},t) = e\varphi_0 \exp(i\mathbf{q}\mathbf{r} - i\omega t) \qquad (1.2)$$

which leads to the appearance of an addition to the off-diagonal components of the density matrix:

$$\rho_{k,k-q} = \frac{e\varphi_0 (f_{k-q} - f_k)}{(\hbar\upsilon + \hbar\omega - \varepsilon_k + \varepsilon_{k-q})} \exp(-i\omega t) \qquad (1.3)$$

where $f_k = f(k) = \rho_{k,k}$ is the electron distribution function, $\varepsilon_k = \varepsilon(k)$ is the electron energy. Eq. (1.3) can be obtained from the equation for off diagonal element of the density matrix [22]:

$$i\hbar\dot{\rho}_{k,k-q} = [H\rho]_{k,k-q} - i\hbar\rho_{k,k-q}\upsilon \qquad (1.4)$$

where $\upsilon$ is the phase relaxation frequency.

The induced charge density is:

$$\sigma = \frac{2e}{S} \sum_k \rho_{k,k-q} \exp(iqr) =$$
$$= \frac{2e^2}{S} \sum_k \frac{\varphi_0 (f_{k-q} - f_k)}{(\hbar \upsilon + \hbar \omega - \varepsilon_k + \varepsilon_{k-q})} \qquad (1.5)$$

where $S$ is the square of a quantum well. In (1.5) we take in to account spin degeneracy. The polarization satisfies the following relation $-div\mathbf{P} = \sigma$, from which in the considered case we obtain $\mathbf{P} = i\mathbf{q}\sigma / q^2$. On the other hand, the potential is related to the magnitude of the electric field: $\mathbf{E} = -\nabla \varphi = -i\mathbf{q}\varphi$. Using these relations, we find the relationship between the polarization and the electric field:

$$\mathbf{P} = -\frac{2e^2 \mathbf{E}}{q^2 S} \sum_k \frac{(f_{k-q} - f_k)}{(i\hbar \upsilon + \hbar \omega - \varepsilon_k + \varepsilon_{k-q})} = \chi \mathbf{E} \qquad (1.6)$$

From (1. 6) we find expression for electron susceptibility

$$\chi = \frac{2e^2}{q^2 S} \sum_k \frac{(f_k - f_{k-q})}{(i\hbar \upsilon + \hbar \omega - \varepsilon_k + \varepsilon_{k-q})} =$$
$$= \frac{2e^2}{q^2 S} \sum_k \frac{(f_k - f_{k+q})}{(\varepsilon_{k+q} - \varepsilon_k - \hbar \omega - i\hbar \upsilon)} \qquad (1.7)$$

that corresponds with Eq. (3) of paper. If $\upsilon \to 0$ then (1.7) becomes Lindhard formula for 2D electron gas [21].

## Appendix 2
## Plasmon quantization

The energy of the electromagnetic field in a dispersive medium is (L.D.Landau, E.M. Lifshchits, Electrodynamics of Continuous Media, v. 8 of Theoretical Physics , A. Vagov, et al, Phys. Rev. B 93, 195414 (2016))

$$H_e = \frac{1}{16\pi} \int d^3 x \left[ \mathbf{E}_q^* \frac{d\omega \bar{\varepsilon}}{d\omega} \mathbf{E}_q + H_q^* H_q \right]. \qquad (2.1)$$

In the quasi-static approximation, when the speed of the plasmon is much less than the speed of light, the last term makes a small contribution, so it can be neglected ($H \sim E_q c/\omega \ll E$). The dielectric constant can be represented as:

$$\tilde{\varepsilon} = \varepsilon + 4\pi \chi \delta(z), \qquad E \parallel q$$
$$\tilde{\varepsilon} = \varepsilon, \qquad E \perp q \qquad (2.2)$$

where $\varepsilon$ is the dielectric constant of the barriers. Taking these factors into account, (2.1) can be rewritten as

$$H_e = \int \frac{d^3 r}{16\pi} \left\{ \mathbf{E}_{q\parallel}^* \mathbf{E}_{q\parallel} \left[ (\varepsilon + 4\pi \chi \delta(z)) + 4\pi \omega \delta(z) \frac{\partial \chi}{\partial \omega} \right] + \mathbf{E}_{qz}^* \mathbf{E}_{qz} \varepsilon \right\} \qquad (2.3)$$

We represent the plasmon potential in the form:

$$\varphi_q = \left[ a_q \exp(i q \mathbf{r} - i\omega t) + a_q^* \exp(-i q \mathbf{r} + i\omega t) \right] \exp(-q \mid z \mid) \quad (2.4)$$

Then the electric field vector directed along $q$ is equal to

$$\mathbf{E}_{q\parallel} = -i\mathbf{q} \left[ a_q \exp(i q \mathbf{r} - i\omega t) - a_q^* \exp(-i q \mathbf{r} + i\omega t) \right] \exp(-q \mid z \mid) \quad (2.5)$$

For $z$ field component we have:

$$E_{qz} = q \left[ a_q \exp(i q \mathbf{r} - i\omega t) + a_q^* \exp(-i q \mathbf{r} + i\omega t) \right] \exp(-q \mid z \mid) sign(z) \qquad (2.6)$$

Substituting (2.5) and (2.6) into (2.3), we obtain:

$$H_e = \int dx dy dz \frac{q^2 \mid a_q \mid^2 \exp(-2q \mid z \mid)}{8\pi} \times$$
$$\left[ (2\varepsilon + 4\pi \chi \delta(z)) + 4\pi \omega \delta(z) \frac{\partial \chi}{\partial \omega} \right] = \qquad (2.7)$$
$$= \frac{q^2 \mid a_q \mid^2 S}{8\pi} \left[ \left( \frac{2\varepsilon}{q} + 4\pi \chi \right) + 4\pi \omega \frac{\partial \chi}{\partial \omega} \right] = \frac{q^2 \omega \mid a_q \mid^2}{2} S \frac{\partial \chi}{\partial \omega}$$

where $S$ is the area of the structure. The first term in square brackets (2.7) of the expression in the third line is equal to zero, since the plasmon dispersion law is obtained from the expression $\varepsilon + 2\pi q \chi = 0$.

Let's introduce the canonical variables:

$$Q_q = \frac{1}{2} \sqrt{\frac{Sq^2}{\omega} \frac{\partial \chi}{\partial \omega}} \left( a_k + a_k^* \right), \qquad P_q = -\frac{i\omega}{2} \sqrt{\frac{Sq^2}{\omega} \frac{\partial \chi}{\partial \omega}} \left( a_k - a_k^* \right)$$

Then

$$H_e = \frac{1}{2} \left( P_q^2 + \omega^2 Q_q^2 \right) \qquad (2.8)$$

Let us introduce the operators of plasmon creation and annihilation:

$$c_q = \sqrt{\frac{\omega}{2\hbar}} \left( Q_q + \frac{iP_q}{\omega} \right), \; c_q^+ = \sqrt{\frac{\omega}{2\hbar}} \left( Q_q - \frac{iP_q}{\omega} \right) \qquad (2.9)$$

Now we can express $a_q$ in terms of the birth and destruction operators:

$$a_q = \sqrt{\frac{2\hbar}{Sq^2} \left( \frac{\partial \chi}{\partial \omega} \right)^{-1}} c_q, \qquad a_q^* = \sqrt{\frac{2\hbar}{Sq^2} \left( \frac{\partial \chi}{\partial \omega} \right)^{-1}} c_q^+ \quad (2.10)$$

Using the relation $\mathbf{E} = -\frac{\partial \mathbf{A}}{c \partial t}$ for vector potential we find expression for its components lying in the plane of the quantum well:

$$\mathbf{A}_q = -\frac{c\mathbf{q}}{\omega}\sqrt{\frac{2\hbar}{Sq^2}\left(\frac{\partial\chi}{\partial\omega}\right)^{-1}}\left(c_{\mathbf{q}}\exp(i\mathbf{q}\mathbf{r} - i\omega t) + \right.$$

$$\left. c_{\mathbf{q}}^+\exp(-i\mathbf{q}\mathbf{r} + i\omega t)\right) \tag{2.11}$$

that corresponds to Eq. (4) of paper.